# Processing/p5 Defined through Practice and Learning

**Kit Kuksenok**
Processing Foundation

**Lee Tusman**
State University of New York at Purchase

***Abstract*—Processing/p5 libraries across different programming languages enact consistent priorities for creative coding as a designed experience. While different programming language ecosystems, like Java and JavaScript, are each associated with their own affordances, community norms, and patterns of use, Processing/p5 sketches across these languages share similarities. Based on case studies of building an implementation of Processing/p5 in two host languages, JavaScript and Lua, we propose a list of software decision-making guiding aspects that constitute Processing/p5, regardless of host language. We discuss this framework in the context of decisions in other exploratory and creative tools that demonstrate how each of the guiding aspects can be operationalized differently than in the case studies. The proposed list highlights opportunities for learning, research, and artistic practice through creation of new Processing/p5 libraries for creative coding and algorithmic art.**

■ **PROCESSING** (Java) and p5.js (JavaScript) are some of the most common open-source tools that people use to learn and make algorithmic art, also known as generative art. Users of these tools can write short programs to create drawings, sound, and interactive artworks. Beginners need no programming knowledge to create artworks with a few lines of code. Artists and educators can also develop design systems and community tools. Although Java and JavaScript are different, Processing (Java) and p5.js (JavaScript) share intentional commonalities in practice and for learners. This paper introduces a definition to refer to these commonalities, and proposes five priorities that can guide the design of creative coding experience.

After nearly 25 years since the release of Processing Java, and over 12 years since development began on p5.js, the ecosystem around these two projects includes around 300 add-on libraries, over five thousand forks, and millions of users worldwide. Both are creative coding tools that imagine "what it would look like for code to become both a creative medium and part of the creative process itself" [1]. These Processing/p5 implementations have been co-created with hundreds of contributors and designed for expressivity and learning, inclusive community [2], and offering “a unique level of variety and control” to artists [3]. However, what, exactly, is Processing/p5?

As one example, processing.py gives a holistic definition: “Processing is not a single programming language, but an arts-centric system for learning, teaching, and making visual form with code” [4]. Aside from this “Python mode” of Processing, there are numerous other implementations of Processing/p5 or libraries that translate a familiar set of tools and sensibilities across other languages. In this essay, we use "Processing/p5" intentionally without reference to specific host language to describe commonly-held technical and community values of many of these projects. Based on reflections on the history of Processing/p5 in JavaScript (including p5.js, `https://p5js.org`) and the recent creation of Processing/p5 in Lua (L5, `https://l5lua.org`), we articulate five aspects that guide the design of a creative coding experience:

- *Situatedness of development environment*—The tool is designed to meet users in their existing context: provide accessible entry points (dedicated software, web editors and bundled environments) and support the devices users already have and are familiar with.
- *Documentation for access and inclusion*—Can include beginner-friendly, little-to-no-prerequisites content, informed by learners and teachers; localization of both reference documentation and error messages; and intentional prioritization of other dimensions of access.
- *Iteration & exploration*—Prioritized through intuitive defaults, open-ended execution using the `setup()/draw()` sketch structure, minimal boilerplate, example driven learning, and immediate visual feedback.
- *Sequential instructions*—Sketch code can be written and read in a linear progression with observable accumulation of graphical state, such as a queue to push to / pop from, and imperative commands preferred to objects.
- *Learning from debugging*—Through the Friendly Error System (FES) as well as naming and parameterization style, API design choices attempt to anticipate common mistakes by beginners, and orient error behavior to support learning where possible.

Creative coding as a designed experience is informed by programming language together with the development environment and technical documentation. Processing Java and p5.js are libraries for sketching with code [5] that introduce expressive consistency across host languages that, respectively, are associated with distinct developer experiences. Each aspect in the proposed framework suggests priorities to guide decision making and community engagement in design and development of a language, library, or the attendant tools, like reference documentation or development environment. In our discussion, we use examples of design and software choices in other coding environments, including Hydra, Arduino, Twine, and Jupyter Notebook, to demonstrate how similar priorities can be addressed in different ways from the choices described by the case studies.

## RELATED WORK

What role do creative coding tools play in how people make, and learn to make, algorithmic art? We highlight several studies on creative coding tools, programming education and software engineering practices in art. This selection informs understanding the effects of tools on learning and practice.

Visual artists use code to make artwork as well as to develop tools [6]. Code allows the creation of their own tools that, while potentially constrained in functionality, provide creative freedom. Through building and using artist-developed software tools, including toolkits and libraries, they engage in community building. As "artists move from creating artifacts to authoring software," usage "is shaped by interactions with technical communities" [ibid.]. Artists working in code may also describe a "practice of designing the goal at the same time as experimenting in code" [7].

Exploratory coding practice, or "code sketching," includes artists and designers rapidly iterating through creative variations. Analogous to traditional sketching, practitioners may generate, discard, or adapt many versions while working through ideas in a "software sketchbook." [5] Processing/p5 languages are notable for supporting this approach.

The Processing/p5 ecosystem includes both the tools themselves, and many tutorials that emphasize sharing and remixing code, and a large collection of public sketches created by other learners. In a study of 1.2 million sketches on OpenProcessing, an online platform for creating and sharing generative art in p5.js, 30% were remixed from other sketches; from those, more than half (55.3%) were "tuning" remixes made by "editing pre-existing parameters" [8]. These remixing practices suggest considering existing community materials, and the design of the sharing platforms around them, as a part of understanding creative coding as a designed experience.

Research on Processing/p5 code editors shows how editor design impacts learning. One study found that "slowness" in code editing proved valuable for "production of and critical engagement with art," while speed and automation enabled rapid exploration [9]. However, McNutt et al. caution that "overeager evaluation can overwhelm and stress users through distracting updates that are unsynchronized with their expected edit-run cadence" [10]. While live feedback is valuable, updates must align with users' work rhythm without interrupting thought. Creative coding environments require consideration of when and how feedback appears.

Chung and Guo's interviews with artist-educators reveal that teaching computing in artistic contexts prioritizes cultural transformation over technical knowledge transfer. Workshops were designed "to prepare participants to create critical interpretations of computing outside of mainstream tech career pipelines" [11]. Teaching and learning code weave together in artistic livelihood and practice, expanding beyond formal education settings. Although the focus of Processing/p5 is learning, this does not mean that this constrains the use and practice of Processing/p5 to education, or to informal learning practices separate from artistic practice.

In a study of a new Educational Programming Language (EPL) and the associated editor, tools, and community governance model, Ko et al. [12] examine accessibility and multilingual support through the design and evaluation of the Wordplay platform, designed for multimodal input (typing, speech, gaze, and switch inputs) and as a "a translanguaging space in which youth are affirmed for their creative and constructive use of multiple language assets." They report that this approach to accessibility and language-inclusive design shows promise, but that numerous design challenges remain, including creating flexibility while maintaining an understandable scope in the design process of a coding experience for learning.

Prior work has established how creative coding tools act as mediators of algorithmic art learning and practice. Research on creative coding tools affordances (liveness) and design (accessibility and governance) inform our case study synthesis through empirical observations of novice engagement, skill development, and sustained artistic practice.

## METHODS

We first present two case studies separately presented as technical accounts of specific design choices. The "Synthesis" section draws on them together to develop the five aspects that guide the design of a creative coding experience, from our perspective as the respective project leads for p5.js and L5. The "Discussion" section then uses this framework to find other operationalizations of these priorities in other software, which be brought back into a Processing/p5 implementation, including p5.js and L5.

The two cases both reflect the design and implementation of libraries with similar technical, artistic, and community goals of inclusion and access, subject to distinct technical constraints, especially resource use. The history of Processing/p5 in JavaScript spans over a decade, while L5 is a new implementation. Although p5.js was intended to be minimal and performant on older hardware, its focus on web browser portability meant it has grown alongside developments in web programming generally. While L5 is not accessible as widely as a browser-based library, it engages with the same current and historical intent that p5.js does.

By highlighting aspects of each implementation that intentionally derive from Processing/p5 into a new host language, these case studies inform our framework for creating new Processing/p5 implementations. For half a year, since the first month of L5's inception, the authors met monthly to discuss developments in L5, to both theoretically and practically understand what is needed for something to be a Processing/p5 implementation. Our in-progress work, including this manuscript, was shared with Processing Foundation team and past and current maintainers of Processing/p5 software. The framework developed during this research also became useful for technical and non-technical discussions of new p5.js features, particularly recent work on JavaScript shader authoring [13].

## CASE STUDIES

We trace the histories of Processing/p5 in two different programming languages, JavaScript and Lua, and integrate them into a framework of shared prioritizations in the "Synthesis" section. We include p5.js, a mature implementation with over a decade of use and contributions by over 800 contributors. We also include L5, an emerging implementation motivated by a lack of Processing/p5 in the Lua language and by goals of creating a creative coding tool that works on a variety of hardware, including older computers.

## Processing/p5 in JavaScript

The first port of Processing into JavaScript was Processing.js, initiated by John Resig in 2008 and maintained until 2016. Prioritizing feature parity with Processing, it allowed existing Processing programs to run, translating their code to JavaScript. Because Java blocks execution during loading, while JavaScript is asynchronous, the structure of Processing.js sketches included an optional `preload()` function, executed prior to `setup()`, for loading any media. In 2013, Lauren Lee McCarthy initiated p5.js as a native port of the ideas of Processing in JavaScript. Unlike Processing.js, it didn't emphasize exact feature parity but instead reimagined Processing for the web.

Though creating p5.js sketches is possible in different environments, the p5.js Editor provides a lightweight, accessible, and widely-used web-based starting point that allows getting started with code immediately, drawing from community-maintained sketches, tutorials, and library ecosystem. Historically, Processing in Java (`https://processing.org`) has been interconnected with the Processing Development Environment (PDE) and community resources such as its documentation and focus on learning materials as a central aspect of Processing's own definition and stated Core Values (`processing.org/overview`).

From its inception, p5.js was created as an entry point for women and people of color to open source development in a field that was lacking in representation [2]. In 2019, the p5.js contributor community adopted the philosophy to only consider features that increase inclusion and access across multiple dimensions, including beginner-friendliness and a focus on historically marginalized communities [14]. Whether an open-source project, like p5.js, is accessible to someone has both an individual and community-level meaning. At an individual level, it may mean understanding and copying the code "quickly and easily", while at a community level, it requires "entry points for people coming from a wide range of abilities and perspectives" into the co-creation process [15]. For example, in the case of web accessibility for screen-readers in p5.js, efforts have included both language features that sketch authors can use, like `describe(…)` and `describeElement(…)`, to address the accessibility limitations of the HTML canvas [16], and labels in the reference website as well as on the p5.js Editor.

Design and development of Processing/p5 integrates educator community feedback through working directly with educators. For example, over the past two years the Art + Code professional development (PD) program that has supported 54 middle and high school teachers to learn to teach with p5.js. Of these, 44 filled out a post-survey. The teachers taught subjects outside of computing (typically, Art), and over 68% of survey respondents responded that they did not "feel like a coder" at all before the PD. In addition to understanding "beginner-friendliness" as being quick to pick up, p5.js and its ecosystem of education materials also supports educators across disciplines to effectively transition from learners of p5.js to teaching creative coding to their own students. 93% of the survey respondents felt more "like a coder" after the PD, based on a self-reported 5-point scale. Although data is not yet available, some have been integrating p5.js into their teaching. During the PD, teachers worked with a recent release of p5.js, and they asked questions and provided candid feedback that directly fed into discussions at Processing Foundation and among p5.js maintainers about how the editor and the documentation can better support teachers. This kind of program provides a valuable input to the design and improvement of p5.js.

## Processing/p5 in Lua

The creation of the L5 library was motivated by teaching in Lua, a language not covered by existing libraries in Processing/p5. In this case study, we consider the implications of referring to L5 as a Processing/p5 library, in terms of API design as well as community activation.

The structure and language of L5 establishes it as Processing/p5, rather than a more general creative coding library. Coders familiar with Processing or p5.js can apply their knowledge directly: typing `rect(x, y, w, h)` or `fill(255, 0, 0)` in L5 works as expected, as does the familiar `setup()` and `draw()` structure that separates initialization from continuous rendering, a pedagogical choice for introducing programming through visual feedback. L5 replicates Processing/p5's approach across core categories (graphics primitives, styling functions, coordinate systems, and color handling) maintaining similar names, parameter order, and behaviors as they would appear in Processing for a subset of methods and variables. These consistencies, combined with sensible defaults, enable sketch portability and allow users familiar with Processing/p5 to begin working in L5 immediately.

Beyond providing specific, familiar building blocks for algorithmic art, adopting Processing/p5 idioms enables iterative sketching in code as the fundamental creative gesture. Sketching with code can mean "freely and quickly trying out a large number of variations of a loosely defined idea [until] a more concrete conception of what is desired takes form," with the expectation that many sketches are not kept or reused [5]. Processing (Java) pioneered programming environments that prioritize immediacy and expressiveness over software engineering concerns [1], and L5 provides, through simple declarative graphics, functions mirroring p5.js's evolution of Processing's original vision.

L5 replicates Processing/p5 functionality, adding to Processing idioms a prioritization of minimal dependencies, low resource consumption, and long-term stability on older hardware. This manifests in deliberate technical choices: L5 currently omits 3D drawing capabilities (as did p5.js 0.x) to ensure operation on underpowered devices, and development centers on debugging core functionality for continued fidelity across computers and older machines rather than expanding the feature set. The documentation reflects this commitment. It can be downloaded for offline use, with options for download without images, and online pages served not to exceed 700KB. This resource dimension articulates not a departure from Processing/p5, but a response to similar intents that p5.js addressed when it was initiated in 2013. Today, L5 arrives at a different conclusion about balancing web accessibility with minimal resource usage. Design for access is a core component of p5.js implementation; L5 investigates what creative coding tools look like when resource constraints become a primary access concern while maintaining a beginner-friendly entry point through Processing/p5 legibility.

L5's licensing and contribution model follow Processing/p5 practices. Licensed under LGPL 2.1, L5 adopts the All Contributors specification [18], recognizing that valuable contributions include documentation, design, community support, and accessibility work alongside code. The emphasis on inline documentation and beginner-friendly explanations instantiates Processing/p5's pedagogical approach, prioritizing accessibility over feature maximalism. As an example of the Processing/p5 community-oriented approach in practice, current community discussion around shader fallbacks for older hardware exemplify that technical decisions happen through open dialogue rather than top-down specification, balancing performance with L5's commitments to longevity and device compatibility in ways that echo Processing/p5's access statement [14].

## SYNTHESIS

We synthesize our reflections on the development of p5.js and L5 into five aspects that guide the design of a Processing/p5 creative coding experience. Listed in order from more contextual, to more code-oriented, these reflect a particular prioritization but are not unique to Processing/p5; examples of other operationalizations, based on existing creative or exploratory programming tools, are provided in the "Discussion" section. Additional to high-level description, this section includes a short illustration of what the alternative to the prioritization for each aspect might look like. The listed alternatives are both reasonable design goals in artist tool development, and not always mutually exclusive with each prioritized aspect. Not everything can be prioritized at once. This section articulates the consistent prioritizations apparent in recent and historical development choices in p5.js and L5, as summarized in "Case Studies".

### Situatedness of Development Environment

The design of Processing/p5 intentionally situates the language and its development environment in the material, social and embodied context of its users. This includes care of both dedicated or bundled software where most learners would first encounter the language, and the devices that this software supports, such as screen-reader support creating entry points for a wider range of users and contributors [15]. For example, p5.js's `describe(…)` method [16] constitutes an essential language feature and a form of situating Processing/p5 in the user's context, not only in principle, but also through research situated in that context. The development of p5.js's `describe(...)` began in 2016 and integrated workshops, focus groups, and feedback from blind and low-vision users [17]. L5 continues this practice of situated design by focusing on users excluded from contemporary creative coding tools, languages and libraries due to limited hardware capabilities or consistent internet access.

*Alternative prioritization:* top-down technical roadmap specific to a particular artistic medium or computational capability, with clear intentional expressive limitations, or focused on library only without emphasis on bundled development tooling.

## Documentation

Documentation functions as a core component of the codebase. One implication of this is that bugfixing encompasses both code updates and clarifying revisions to documentation. In an open-source project, issues and bug reports can function to highlight these gaps. This approach recognizes that bugs can originate in the gap between the API designer's intent and the expectation of the user (or learner). Addressing technical errors includes understanding and designing bridges and mediations for this gap. In addition to the reference materials, the Friendly Error System (FES) of p5.js also has support for localization/internationalization, with translation strings available in users' languages. L5's development process similarly uses feedback to identify and address user expectations. Compared to Processing's dedicated IDE and the p5.js web editor, L5 exposes more seams: users must install the Love2d dependency and bring their own code editor.

Documentation in Processing/p5 also challenges common expectations of API designers about what really needs to be a prerequisite. As a tool, p5.js can be described as both "low-floor" (allowing beginners to produce visual or interactive output quickly, often within minutes of beginning to learn the language) and "high ceiling" (p5.js is hosted in JavaScript, and a learner can access not only the functionality of many p5.js community add-on libraries, but also other web browser capabilities as they develop their practice) [19]. The "low-floor" aspect of p5.js is not created by constraining what is possible, but by creating entry points that require as little prior knowledge as possible.

For example, drawing a circle: the Mozilla Canvas API uses the generalized `arc(…)` function, requiring some trigonometry knowledge [20], whereas p5.js provides a dedicated `circle(…)` function with a documentation that includes a short definition of radius, requiring no prior geometry knowledge [21]. Beyond this low floor for drawing basic shapes, p5.js raises the ceiling through generalizations like `ellipseMode(…)`, which applies to circles, arcs, and ellipses, making explicit what the Canvas API [20] holds as prerequisite knowledge. Furthermore, the abstractions of p5.js work across renderers (Canvas, WebGL, WebGPU and the community-maintained SVG renderer p5.js-svg) with minimal syntax changes, raising the ceiling through the interplay of API design and documentation.

*Alternative prioritization:* technically dense documentation that builds on domain expertise.

## Iteration & Exploration

The Processing/p5 sketch structure centers exploratory continuous exploration rather than focusing on task completion. When the sketch runs, the `setup()/draw()` structure separates code executed once in the beginning, from code executed in an ongoing loop. Code can also be triggered by event handlers such as `mousePressed()` or `keyPressed()`, which typically also allow drawing operations outside of the draw loop. Because code continues to execute frame by frame until stopped or terminated from the runtime, programmers observe their sketches continuously and indefinitely, with no termination values or conditions. In development environments that enable auto-refresh (including the p5.js Editor), this runtime iteration allows design iteration: to continuously modify the code and observe the resulting dynamics.

The Processing/p5 API design, in terms of function naming and the wide use of intuitive defaults and optional parameters, also enables design iteration. API decisions explicitly minimize the changes needed to explore visually meaningful changes: `circle(x, y, diameter)` does not require objects, color, stroke or context-lowering the threshold for producing visual output, while preserving access to affordances for fine-grained control. This architecture supports exploratory and improvisational coding practices within the concept of Processing's original vision of a "digital sketchbook" for "sketching" with code [1], a practice distinct from both rapid prototyping and systematic software engineering. Where prototyping emphasizes intentional progression toward a refined end goal, Processing/p5 encourages open-ended experimental play and discovery as common computational gestures. Code becomes material for continuous transformation and remixing [8]. The code sketch can exist both as finished work and a perpetual draft, collapsing distinctions between authorship and adaptation, completion and continuation.

*Alternative prioritization:* Library that prioritizes writing code that is reusable, efficient and reliable, over ease of iterative editing of behavior, such as through explicit and verbose parameterization.

## Sequential Instructions

Processing/p5 privileges sequential composition as a foundational organizational principle. Drawing instructions accumulate effects in a stateful queue without requiring object manipulation: `fill(255, 0, 0)` modifies the global rendering of color fill such that following shapes use this color until explicitly changed. Commands like `rect()`, `ellipse()` and `line()` function as imperative, non-object-oriented commands creating both linear execution flow, and a legible sequence of drawing actions. Sequential composition extends to transformation matrices (including `translate()`, `rotate()`, `scale()`) which modify the coordinate system successively, while `push()/pop()` enables hierarchical composition and preserve sequential layer ordering in Processing, p5.js and L5. Unlike an object-oriented graphics framework employing scene graphs or retained rendering, Processing/p5 sketches contain code that is written, read, and often executed in a linear progression with visible accumulation of the graphical state. Though some retained capabilities (Processing's `PShape` or `p5.Geometry`) exist, they are examples of opt-in uses that raise the ceiling beyond sequential gestures.

*Alternative prioritization:* A library that privileges modular abstraction and composition of user-defined functions, so even trying to create top-to-bottom reading order would be difficult. Some high-ceiling functionality in p5.js exhibits this in practice. For example, a JavaScript shader authoring feature in p5.js, which does not use "sequential" execution or data flow, community feedback resulted in a "flatter" sequential composition in some parts of the related API, though not all of it [13].

## Learning from Debugging

API design choices attempt to anticipate common mistakes by beginners, and orient error behavior to support learning where possible. We have outlined various ways that the functionality of Processing/p5 can be familiar across different host programming languages, but there are also differences depending on the affordances of the programming languages that the libraries are based on. Java and L5 have compilation errors, but JavaScript does not. p5.js's Friendly Error System (FES) intercepts and interprets the errors, providing non-blocking explanations, to support the user to recover from the error while learning by doing, and is an example of a feature whose basic behavior and implementation must be informed by the host language, even if drawing from a common Processing/p5 sensibility and design approach. For example, the p5.js `circle(...)` implementation expects at least 3 arguments [20]. While JavaScript doesn't enforce parameter counts, FES provides immediate feedback ("circle() expected 3 arguments, received 2"), preventing cryptic downstream errors, silent failures, or TypeScript requirements.

From the perspective of the artist, coder, or learner, both iterative code sketching and sequential instructions are continuous processes interrupted by bugs, errors, and unexpected behavior. These interruptions become teaching moments when bugs are treated as inevitable learning opportunities rather than failures to avoid. This implies that intuitive defaults should not create complexity traps that exceed the coder's debugging ability; and that debugging should guide learners forward to new use rather than prevent misuse.

The `circle(…)` example illustrates this: learners create circles without trigonometry concepts, incorporate color and animation before understanding objects or loops. p5.js provides artistic expression with no computer science prerequisites and intentionally designed error behavior. This allows interleaving learning artistic concepts and computational thinking concepts. Apparent in both recent Art + Code PD and in Processing's history (`processing.org/overview`), Processing/p5 is stress-tested through years of use by teachers who are themselves novice coders learning not only how to code, but also how to teach computational thinking.

*Speculative counterexample:* error message style that prioritizes verbosity or provides feedback on fixing the problem, but not necessarily on what should be learned next.

# DISCUSSION

The five Processing/p5 guiding aspects for software design for creative coding can be operationalized in different ways than those that p5.js and L5 have chosen. In this section, we draw from an established ecosystem outside of Processing/p5. Each of Hydra, Arduino, Twine, and Jupyter Notebook demonstrates a design choice that operationalizes one of the five aspects in a way that is distinct from what we have presented in p5.js and L5. Developers of either Processing/p5 or other creative or exploratory programming tools can use this framework to find commonalities and differences between what makes existing tools accessible for learning and creative practice, even if those tools were not using this framework during their own development.

Hydra supports learning through beginner-friendly “getting started” features, including starting with community-written programs and visual tools for exploring the parameter space of these programs. Hydra, is a “live code-able video synth and coding environment that runs directly in the browser,” taking inspiration from “analog modular synthesis, in which chaining or patching a set of transformations together generates a visual result” [22]. Coding using Hydra is situated in a lightweight, easy to use web-based editor that invites sharing, collaboration, and remixing: even after start-up, authors can randomize and go to another community sketch, or randomize parameters in the current sketch. These features allow exploration & iteration not only through code but also through immediate direct engagement with community work.

Arduino supports learning through a well-organized tutorial catalogue browser. Arduino is an “open-source electronics platform” that enables a user to “tell [a] board what to do by sending a set of instructions to the microcontroller on the board” [23]. Development is situated both within a particular physical device, and a web-based alternative to downloadable software. The tutorial catalogue browser uses labels to make navigating a wide landscape of heterogeneous devices more accessible to beginners.

Twine supports learning and creative exploration through debugging tools that make complex program state visible without requiring knowledge of web developer tooling. Twine is “an open-source tool for telling interactive, non-linear stories” [24]. To enable iteration & exploration, the editor combines visual and textual programming authorship features and allows the user to execute from any point in their interactive story. The web-based testing tool enables debugging using navigation and variable inspection that would be hidden in a running story.

Jupyter Notebook supports creative exploration through enabling package management within a consistent iterative workflow. Machine learning and data analysis has been of increased interest to creative coders and can include work with interactive Python code in tools like Jupyter Notebook. Development can be situated in a variety of web-based options that are sharable. Code cells in a notebook are not limited to Python, and can also run command line calls, including recovering from missing dependency bugs by running the installation directly in the environment.

The tools each operationalize one of the guiding aspects, such as: Hydra operationalizes situatedness through community-sketch startup, and iteration & exploration through parameter randomization. Arduino operationalizes beginner-friendly documentation through its hierarchical tutorial catalogue. Both Twine and Jupyter Notebook operationalize learning from debugging in two very different ways: Twine by surfacing program state inspection without developer tooling, and Jupyter Notebook through in-environment package management that turns dependency errors into recoverable steps.

## CONCLUSION

In this essay, we articulate an initial definition of Processing/p5 through a set of five aspects that guide the design of a creative coding library. Creating new variants of Processing/p5 helps artists or educators leverage existing Processing/p5 fluency in new host languages, different technical ecosystems, or in response to priorities or constraints not previously explored by existing Processing/p5 implementations. The L5 case study exemplifies how these aspects can guide implementation while navigating resource constraints, pedagogical priorities, and requirements for inclusion and access.

The proposed framework is the outcome of iterative, top-down synthesis by the authors based on our experiences working on p5.js and L5, and reading of other established projects. Although the projects considered are built in community, the framework expresses a maintainer/contributor perspective. Future research should interrogate this framework through bottom-up empirical study, including interviews or surveys of creative coding novices, educators, and practitioners, to understand how users or learners understand the experience of engaging with Processing/p5 from different entry points and with different access needs. Qualitative and quantitative study of learning during and through debugging could also be valuable contributions not only to creative coding, but to software engineering education research more generally. Additionally, autoethnographic reflections of novices creating their own Processing/p5 variants could explore mechanisms of API design as part of creative coding education. Finally, the framework provides a high-level lens for tool comparison. More in-depth evaluations and recommendations for each of the five aspects would not only help understand and improve existing or emerging Processing/p5 libraries, but also the creative coding environments that historically or incidentally share practices or user-bases with Processing/p5.

The history of Processing/p5 is associated with a particular visual aesthetic. Multiple other styles and modalities have been explored, including interactive sound art, XR projects, and performance or theater applications. The proposed framework be applied to non-visual and not-only-visual creative media. We hope that new and experimental Processing/p5 libraries can explore shifts and expansions of this creative medium, while providing a familiar, inclusive, and accessible entry point for learners and practitioners.

## ACKNOWLEDGMENT

We thank Raphaël de Courville, Roxana Hadad, Dave Pagurek van Mossel, Cassie Tarakajian, and Amy B. Woodman for their input and feedback. We also thank the reviewers for their supportive and thoughtful feedback.

## REFERENCES

1. L. Stinson, "Processing: The software that shaped creative coding," Eye on Design, AIGA, Oct. 28, 2021.Oral history with B. Fry, C. Reas, and D. Shiffman. [Online]. Available: https://eyeondesign.aiga.org/processing-the-software-that-shaped-creative-coding/. [Accessed: Dec. 15, 2025].
2 L. Tusman, L. L. McCarthy, D. R. Santos, M. Turner, C. Tarakajian. "Can a Programming Language be a Radical Community?," Artists and Hackers, Feb. 26, 2021. [Online]. Available: https://www.artistsandhackers.org/programming-radical-community . [Accessed: Dec. 15, 2025].
3. D., Pagurek van Mossel, K. Kuksenok, "Greetings from p5.js 2.0: Animation, Interaction, and Typography in 2D and 3D," In Proceedings of the Special Interest Group on Computer Graphics and Interactive Techniques Conference (SIGGRAPH) Labs (pp. 1-2), August 2025.
4 J. Feinberg, "Processing.py," 2022. [Online] Available: https://py.processing.org/reference/ [Accessed: Dec. 15, 2025].
5. I. Bergström, A.F. Blackwell. (2016, September). "The practices of programming." In 2016 IEE symposium on visual languages and human-centric computing (VL/HCC) (pp. 190-198). IEEE.
6. J.,Li, S., Hashim, & J. Jacobs (2021, May). "What we can learn from visual artists about software development." In Proceedings of the 2021 CHI Conference on Human Factors in Computing Systems (pp. 1-14).
7. M.B., Kery, A.B., Myers. (2017, October). "Exploring exploratory programming." In 2017 IEEE Symposium on Visual Languages and Human-Centric Computing (VL/HCC) (pp. 25-29). IEEE.
8. B. Subbaraman, S. Shim, N. Peek. "Forking a sketch: How the OpenProcessing community uses remixing to collect, annotate, tune, and extend creative code." Proceedings of the 2023 ACM Designing Interactive Systems (DIS) Conference (pp. 326-342), 2023.
9. A.M. McNutt, Cohen, S. and Chugh, R., "Slowness, Politics, and Joy: Values That Guide Technology Choices in Creative Coding Classrooms." In Proceedings of the 2025 CHI Conference on Human Factors in Computing Systems (pp. 1-16). April, 2025.
10. A.M., Mcnutt, A. M., Outkine, A., & Chugh, R. (2023, April). A study of editor features in a creative coding classroom. In Proceedings of the 2023 CHI Conference on Human Factors in Computing Systems (pp. 1-15).
11. A. M. Chung, P. J. Guo (2024, August). Perpetual teaching across temporary places: Conditions, motivations, and practices of media artists teaching computing workshops. In Proceedings of the 2024 ACM Conference on International Computing Education Research-Volume 1 (pp. 374-388).
12. A. J. Ko, Aldana Lira, C., & Amaya, I. (2025, April). Wordplay: Accessible, Multilingual, Interactive Typography. In Proceedings of the 2025 CHI Conference on Human Factors in Computing Systems (pp. 1-20).
13. D. Pagurek, L. Plowden, P. Singh, K. Lim, K. Kuksenok. [Online] "Beginner-Friendly Shader Programming in p5.js v2 -- talk." Libre Graphics Meeting 2026 Available: https://media.ccc.de/v/lgm-2026-110664-beginner-friendly-shader-programming-in-p5-js-v2 [Accessed: May 18, 2026].
14. p5.js contributors, "Our Focus on Access," 2019. [Online]. Available: https://p5js.org/contribute/access/ [Accessed: Dec. 15, 2025]
15. R. Vasudevan, "Aligning an Open Source Ethos," 2024. [Online] Available: https://www.opensourceethos.net/ [Accessed: Dec. 15, 2025].
16. K. Lichlyter, L. Quiñones, "Writing Accessible Canvas Descriptions." [Online] Available: https://p5js.org/tutorials/writing-accessible-canvasdescriptions/ [Accessed: Dec. 15, 2025].
17. C. Kearney-Volpe, "P5 Accessibility," Dec 1, 2017. [Online] Available: https://medium.com/processingfoundation/p5-accessibility-115d84535fa8 [Accessed: Dec. 15, 2025].
18. "All Contributors" [Online] Available: https://allcontributors.org/ [Accessed: Dec. 15, 2025].
19. S. Papert, "Mindstorms: Children, computers, and powerful ideas," 1980, Basic Books. as referenced by J.C. Blake-West and M. U. Bers. "ScratchJr design in practice: Low floor, high ceiling." International Journal of Child-Computer Interaction 37 (2023): 100601.
20. mdn, "Drawing shapes with canvas" [Online] Available: https://developer.mozilla.org/en-US/docs/Web/API/Canvas_API/Tutorial/Drawing_shapsh#arcs [Accessed: Dec. 15, 2025].
21. p5.js contributors, "circle." [Online] Available: https://p5js.org/reference/p5/circle/
22. O. Jack, "hydra live coding video synth," 2025. [Online] Available: https://hydra.ojack.xyz/ [Accessed: Dec. 15, 2025].
23. Arduino, "What is Arduino?," Jan. 2022. [Online] Available: https://docs.arduino.cc/learn/starting-guide/whats-arduino/ [Accessed: Dec. 15, 2025].
24. Interactive Fiction Technology Foundation. "Twine / An open-source tool for telling interactive, nonlinear stories," 2025. [Online] Available: https://twinery.org [Accessed: Dec. 15, 2025].

**Kit Kuksenok** is p5.js Project Lead at Processing Foundation, and a maintainer of the p5.js library. They hold a MSci and PhD in Computer Science and Engineering from the University of Washington. Contact them at kit@processingfoundation.org

**Lee Tuman** is Associate Professor of New Media and Computer Science at SUNY Purchase. He received his MFA in Design Media Arts at University of California, Los Angeles (UCLA). His research interests include generative art, decentralized networks, game engines and language design. He is host of the Artists and Hackers podcast. Contact him at lee.tusman@purchase.edu